\documentclass{acmtrans2m}

\pdfoutput=1 

\usepackage{stfloats}
\usepackage{booktabs}
\usepackage{arydshln}
\usepackage{ifpdf}

\ifpdf
  \usepackage[pdftex]{graphicx}
  \DeclareGraphicsExtensions{.pdf}
\else
  \usepackage{graphicx}
  \DeclareGraphicsExtensions{.ps,.eps,.ai}
\fi

\graphicspath{{tist10-figures/}}

\newcommand{\Ni}{(1)~}
\newcommand{\Nii}{(2)~}
\newcommand{\Niii}{(3)~}

\usepackage{algorithm}
\usepackage[]{algpseudocode}

\usepackage{xspace}
\usepackage{amssymb}
\usepackage{amsmath}
\usepackage{amsfonts}

\newcommand{\real}{\mathbb{R} }

\newcommand{\argmin}{\ensuremath{\mathop{\mathrm{argmin}}}}

\newcommand{\Loss}{\ensuremath{L}\xspace}
\newcommand{\Penalty}{\ensuremath{R}\xspace} 

\newcommand{\sLang}{\ensuremath{\mathcal{S}}\xspace}
\newcommand{\tLang}{\ensuremath{\mathcal{T}}\xspace}

\newcommand{\Classifier}{\ensuremath{f}\xspace}

\newcommand{\sData}{\ensuremath{D_\sLang}\xspace}
\newcommand{\suData}{\ensuremath{D_{\sLang,u}}\xspace}
\newcommand{\tData}{\ensuremath{D_{\tLang}}\xspace} 
\newcommand{\tuData}{\ensuremath{D_{\tLang,u}}\xspace}
\newcommand{\uData}{\ensuremath{D_u}\xspace}

\newcommand{\stMap}{\ensuremath{\theta}\xspace}

\newcommand{\Pivots}{\ensuremath{P}\xspace}

\newcommand{\Voc}{\ensuremath{V}\xspace}
\newcommand{\sVoc}{\ensuremath{V_\sLang}\xspace}
\newcommand{\tVoc}{\ensuremath{V_\tLang}\xspace}
\newcommand{\pVoc}{\ensuremath{V_{P}}\xspace}

\newcommand{\sWord}{\ensuremath{w_\sLang}\xspace}
\newcommand{\tWord}{\ensuremath{w_\tLang}\xspace}

\begin{document}


\markboth{Prettenhofer, Stein}{Cross-Lingual Adaptation using SCL}

\title{Cross-Lingual Adaptation\\ using Structural Correspondence Learning}
\author{PETER PRETTENHOFER\\Bauhaus-Universit{\"a}t Weimar
\and
BENNO STEIN\\Bauhaus-Universit{\"a}t Weimar}

\begin{abstract}
Cross-lingual adaptation, a special case of domain adaptation, refers to the transfer of classification knowledge between two languages. In this article we describe an extension of Structural Correspondence Learning (SCL), a recently proposed algorithm for domain adaptation, for cross-lingual adaptation. The proposed method uses unlabeled documents from both languages, along with a word translation oracle, to induce cross-lingual feature correspondences. From these correspondences a cross-lingual representation is created that enables the transfer of classification knowledge from the source to the target language. The main advantages of this approach over other approaches are its resource efficiency and task specificity. 

We conduct experiments in the area of cross-language topic and sentiment classification involving English as source language and German, French, and Japanese as target languages. The results show a significant improvement of the proposed method over a machine translation baseline, reducing the relative error due to cross-lingual adaptation by an average of 30\% (topic classification) and 59\% (sentiment classification). 
We further report on empirical analyses that reveal insights into the use of unlabeled data, the sensitivity with respect to important hyperparameters, and the nature of the induced cross-lingual correspondences. 
\end{abstract}

[information filtering]

\begin{bottomstuff}
Author's address: P. Prettenhofer, Bauhaus-Universit{\"a}t Weimar, 99421 Weimar, Germany. \newline
\end{bottomstuff}
\maketitle

\section{Introduction}

Over the past two decades supervised machine learning methods have been successfully applied to many problems in natural language processing (e.g., named entity recognition, relation extraction, sentiment analysis) and information retrieval (e.g., text classification, information filtering). These methods, however, rely on large, annotated training corpora, whose acquisition is time-consuming, costly, and inherently language-specific. As a consequence most of the available training corpora are in English only. Since an ever increasing fraction of the textual content available in digital form is written in languages other than English%
\footnote{This is especially the case for the World Wide Web, where from 2000 to 2009 the content available in Chinese grew more than four times as much as the content available in English (\url{http://www.internetworldstats.com/stats7.htm}, June 2010).}
, this limits the widespread application of state-of-the-art techniques from natural language processing (NLP) and information retrieval (IR).
Technology for cross-lingual adaptation aims to overcome this problem by transferring the knowledge encoded within annotated (=~labeled) data written in a source language to create a classifier for a different target language. Cross-lingual adaptation can thus be viewed as a special case of domain adaptation, where each language acts as a separate domain. 

In contrast to ``classical'' domain adaptation, cross-lingual adaptation is characterized by the fact that the two domains, i.e.,\ the languages, have non-overlapping feature spaces, which has both theoretical and practical implications for domain adaptation. In classical domain adaptation---as well as in related problems such as covariate shift---the factor of overlapping feature spaces is implicitly presumed by the following or similar assumptions:
\Ni
generalizable features, i.e., features which behave similarly in both domains, exist \cite{Jiang:2007a,Blitzer:2006,Daume:2007}, or,
\Nii
the support of the test data distribution is contained in the support of the training data distribution \cite{Bickel:2009}. 
If, on the other hand, the feature sets are non-overlapping, one needs external knowledge to link features of the source domain and the target domain \cite{Dai:2008}. 

This article extends the work of \citeN{Prettenhofer:2010} and presents an approach for cross-lingual adaptation in the context of text classification: Cross-Language Structural Correspondence Learning (CL-SCL). 
CL-SCL uses unlabeled data from both languages along with external domain knowledge in the form of a word translation oracle to induce cross-lingual word correspondences. The approach is based on Structural Correspondence Learning (SCL), a recently proposed algorithm for domain adaptation in natural language processing. 

Similar to SCL, CL-SCL induces correspondences among the words from both languages using a small number of so-called {\em pivots}. In CL-SCL, a pivot is a pair of words, $\{\sWord, \tWord\}$, from the source language \sLang and the target language \tLang, which possess a similar semantics. Testing the occurrence of \sWord or \tWord in a set of unlabeled documents from \sLang and \tLang yields two equivalence classes {\em across} these languages: one class contains the documents where either \sWord or \tWord occur, the other class contains the documents where neither \sWord nor \tWord occur. Ideally, a pivot splits the set of unlabeled documents with respect to the semantics that is associated with $\{\sWord, \tWord\}$. The correlation between \sWord or \tWord and other words $w$, $w\not\in \{\sWord, \tWord\}$ is modeled by a linear classifier, which then is used as a language-independent predictor for the two equivalence classes. A small number of pivots can capture a sufficiently large part of the correspondences between \sLang and \tLang in order to
\Ni
construct a cross-lingual representation and
\Nii
learn a classifier that operates on this representation.
Several advantages follow from this approach: 

\begin{longitem}
\item
Task specificity. The approach exploits the words' pragmatics since it considers---during the pivot selection step---task-specific characteristics of language use. \\
\item
Efficiency in terms of linguistic resources. The approach uses unlabeled documents from both languages along with a small number (100 - 500) of translated words, instead of employing a parallel corpus or an extensive bilingual dictionary.
\item
Efficiency in terms of computing resources. The approach solves the classification problem directly, instead of resorting to a more general and potentially much harder problem such as machine translation. 
\end{longitem}

The article is organized as follows: Section~\ref{sec:related-work} discusses cross-lingual adaptation in the context of related work including domain adaptation and dataset shift.
Section~\ref{sec:cross-language-text-classification} introduces the problem of cross-language text classification, a special case of domain adaptation. Section~\ref{sec:cross-language-structural-correspondence-learning} describes Cross-Language Structural Correspondence Learning and proposes a new regularization schema for the pivot predictors. Section~\ref{sec:experiments} reports on the design and the results of experiments in the area of cross-language sentiment and topic classification. Finally, Section~\ref{sec:conclusions} concludes our work.


\section{Related Work}
\label{sec:related-work}

\begin{figure*}
\includegraphics[scale=1.0]{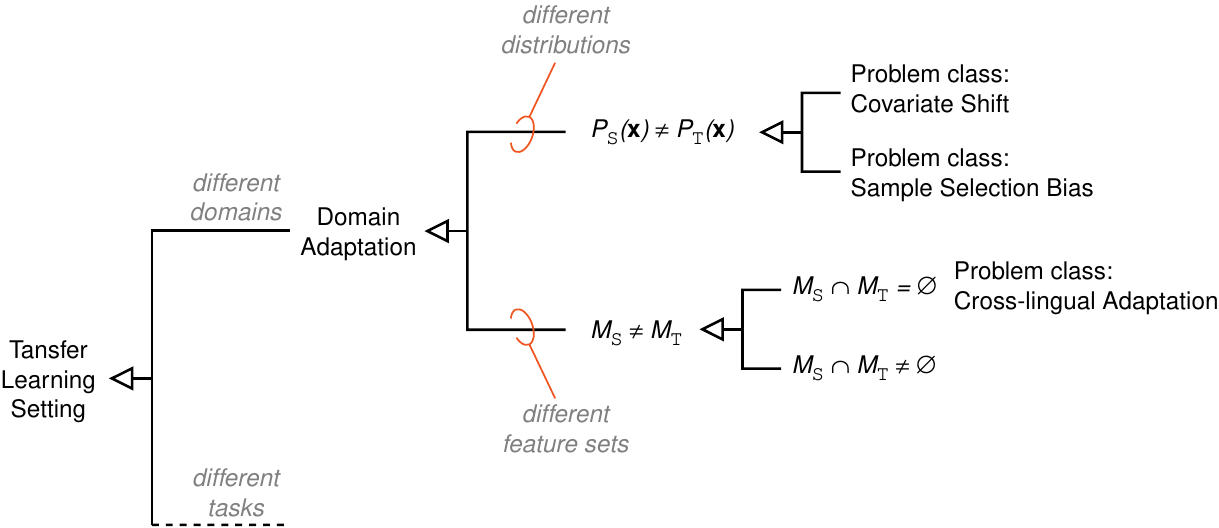}
\caption{A taxonomy of transfer learning settings, organized the dimension ``domain'' and ``task''. The domain adaptation branch is unfolded.}
\label{transfer-learning-taxonomy}
\end{figure*}

The idea to transfer knowledge from a source learning setting $\mathcal{S}$ to a different target learning setting $\mathcal{T}$ is an active field of research \cite{Pan:2009}, and Figure~\ref{transfer-learning-taxonomy} organizes well-known problems within a taxonomy. The taxonomy combines the two most important determinants within a learning setting, namely, the {\em domain} and the {\em task}. A domain is defined by
\Ni
a set of features $M$,
\Nii
a space of possible feature vector realizations $\mathbf{x}$, which typically is the $\mathbb{R}^{|M|}$, and
\Niii
a probability distribution $P(\mathbf{x})$ over the space of possible feature vector realizations.%
\footnote{If clear without ambiguity we use $\mathbf{x}$ or $y$ to denote both a realization and a random variable.}
A task specifies a set of labels corresponding to classes, typically $\{+1,-1\}$, along with a conditional distribution $P(y\mid\mathbf{x})$, with $y\in\{+1,-1\}$. Alternatively, a task can be specified by a sample $\{(\mathbf{x},y)\mid \mathbf{x}\in\mathbb{R}^{|M|}, y\in\{+1,-1\}\}$.
In Figure~\ref{transfer-learning-taxonomy} the domain adaptation branch is unfolded since it is the focus of this article. The branch ``different distributions'' addresses problems where the feature sets are unchanged; without loss of generality $P_\mathcal{S}(\mathbf{x})\not=P_\mathcal{T}(\mathbf{x})$ can also be presumed for problems in the branch ``different feature space''. 


\subsection{Domain Adaptation}

Domain adaptation refers to the problem of adapting a statistical classifier trained on data from one (or more) source domains to a different target domain . In the basic domain adaptation setting we are given labeled data from a source domain $\mathcal{S}$ and unlabeled data from the target domain $\mathcal{T}$, and the goal is to train a classifier for the target domain. Beyond this setting one can further distinguish whether a small amount of labeled data from the target domain is available \cite{Daume:2007,Finkel:2009} or not \cite{Blitzer:2006,Jiang:2007a}. The latter setting is referred to as unsupervised domain adaptation.

\citeN{Blitzer:2006} propose an effective algorithm for unsupervised domain adaptation, called Structural Correspondence Learning. Within a first step SCL identifies features that generalize across domains, which the authors call pivots. SCL then models the correlation between the pivots and all other features by training linear classifiers on the unlabeled data from both domains. This information is used to induce correspondences among features from the different domains and to learn a shared representation that is meaningful across both domains. SCL is related to the structural learning paradigm introduced by \citeN{Ando:2005b}. The basic idea of structural learning is to constrain the hypothesis space of a learning task by considering multiple different but related tasks on the same input space. \citeN{Ando:2005a} present a semi-supervised learning method, Alternating Structural Optimization (ASO), based on this paradigm, which generates related tasks from unlabeled data.
They show that ASO delivers state-of-the-art performance for a variety of natural language processing tasks including named entity and syntactic chunking.
\citeN{Quattoni:2007} apply structural learning to image classification in settings where little labeled data is given. 


\subsection{Dataset Shift}

Traditional machine learning assumes that both training and test examples are drawn from identical distributions. In practice this assumption is often violated, for instance due to the irreproducibility of the test conditions within the training phase. Dataset shift refers to the general problem when the joint distribution of inputs and outputs differs between training phase and test phase. The difference between dataset shift and domain adaptation is subtle; in fact, both refer to the same underlying problem but emerge from the viewpoints of different research communities. Dataset shift is coined by the machine learning community and builds on prior work in statistics, in particular the work on covariate shift \cite{Shimodaira:2000} and sample selection bias \cite{Cortes:2008}. In contrast, domain adaptation originates from the natural language processing community. Most of the early work on domain adaptation focuses on the question of how to leverage ``out-domain data'' (=~data associated with $\mathcal{S}$) effectively to learn a classifier when only little or no labeled ``in-domain data'' (=~data associated with $\mathcal{T}$) is available. The latter emphasizes the relationship to semi-supervised learning---with the crucial difference that labeled and unlabeled data stem from different distributions.
Covariate shift can be considered as a certain case of dataset shift which is closely related to unsupervised domain adaptation. It is characterized by the fact that the class conditional distribution between training phase and test phase is equal, i.e.\ $P_\mathcal{S}(y\mid\mathbf{x}) = P_\mathcal{T}(y\mid\mathbf{x})$, while the marginal distribution of the inputs (covariates) differs, i.e.\ $P_\mathcal{S}(\mathbf{x}) \ne P_\mathcal{T}(\mathbf{x})$.
A broad discussion of dataset shift is beyond the scope of this article; the interested reader is referred to \cite{Quionero-Candela:2009}.


\subsection{Cross-Lingual Adaptation}

Analogous to domain adaptation, cross-lingual adaptation refers to the problem of adapting a statistical classifier trained on data from a source language \sLang to a different target language \tLang. Examples include the adaptation of a named-entity recognizer, a syntactic parser, or a relation extractor. 
The major characteristic of cross-lingual adaptation is the fact that the two "domains" have non-overlapping features sets, i.e., $M_\mathcal{S} \ne M_\mathcal{T}$. 
While cross-lingual adaptation has not received a lot of attention in the  natural language processing community, a special case of cross-lingual adaptation has received a lot of attention recently: cross-language text classification, which is also the focus of this article.  

\citeN{Bel:2003} belong to the first who explicitly considered the problem of cross-language text classification. Their research, however, is predated by work in cross-language information retrieval, CLIR, where similar problems are addressed \cite{Oard:1998}. Traditional approaches to cross-language text classification and CLIR use linguistic resources such as bilingual dictionaries or parallel corpora to induce correspondences between two languages \cite{Lavrenko:2002,Olsson:2005}. \citeN{Dumais:1997a} is considered as seminal work in CLIR: they propose a method which induces semantic correspondences between two languages by performing latent semantic analysis, LSA, on a parallel corpus. \citeN{Li:2007} improve upon this method by employing kernel canonical correlation analysis, CCA, instead of LSA. The major limitation of these approaches is their computational complexity and, in particular, the dependence on a parallel corpus, which is hard to obtain---especially for less resource-rich languages. \citeN{Gliozzo:2005} circumvent the dependence on a parallel corpus by using so-called multilingual domain models, which can be acquired from comparable corpora in an unsupervised manner. In \cite{Gliozzo:2006} they show for particular tasks that their approach can achieve a performance close to that of monolingual text classification. 

Recent work in cross-language text classification focuses on the use of automatic machine translation technology. Most of these methods involve two steps:
\Ni
translation of the documents into the source or the target language, and
\Nii
dimensionality reduction or semi-supervised learning to reduce the noise introduced by the machine translation.
Methods which follow this two-step approach include the EM-based approach by \citeN{Rigutini:2005}, the CCA approach by \citeN{Fortuna:2005}, the information bottleneck approach by \citeN{Ling:2008}, and the co-training approach by \citeN{Wan:2009}.


\section{Cross-Language Text Classification}
\label{sec:cross-language-text-classification}

In standard text classification, a document $d$ is represented under the bag-of-words model as $|\Voc|$-dimensional feature vector $\mathbf{x}\in \mathcal{X}$, where \Voc, the vocabulary, denotes an ordered set of words, $x_i\in\mathbf{x}$ denotes the normalized frequency of word $i$ in $d$, and $\mathcal{X}$ is an inner product space. $\sData$ denotes the training set and comprises tuples of the form $(\mathbf{x}, y)$, which associate a feature vector $\mathbf{x}\in \mathcal{X}$ with a class label $y\in \mathcal{Y}$. For simplicity but without loss of generality we assume binary classification problems, $\mathcal{Y} = \{$+1, -1$\}$. The goal is to find a classifier $\Classifier: \mathcal{X} \to \mathcal{Y}$ that predicts the labels of new, previously unseen documents. In the following, we restrict ourselves to linear classifiers:
\begin{equation}
 \Classifier(\mathbf x) = sign(\mathbf{w}^T \mathbf{x}) ,
\end{equation}

where $\mathbf{w}$ is a weight vector that parameterizes the classifier and $[\cdot]^T$ denotes the matrix transpose. The computation of $\mathbf{w}$ from \sData is referred to as model estimation or training. A common choice for $\mathbf{w}$ is given by a vector $\mathbf{w}^*$ that minimizes the regularized training error:
\begin{equation}
\label{eqn:regularized-training-error}
\mathbf{w}^* = \argmin_{\mathbf{w}\in \real^{|V|}} \kern-2pt \sum_{(\mathbf{x},y) \in\sData} \kern-4pt \Loss(y, \, \mathbf{w}^T\mathbf{x}) + \lambda \Penalty(\mathbf{w}) .
\end{equation}

\Loss is a loss function that measures the quality of the classifier, \Penalty is a regularization term that penalizes model complexity, and $\lambda$ is a non-negative hyperparameter that models the tradeoff between classification performance and model complexity. A common choice for \Penalty is L2-regularization, which imposes an L2-norm penalty on $\mathbf{w}$, $\Penalty(\mathbf{w}) = \frac{1}{2}{\Vert \mathbf{w} \Vert}_2^2 = \frac{1}{2} \, \mathbf{w}^T \mathbf{w}$. Different choices for \Loss entail different classifier types; e.g., when choosing the hinge loss function one obtains the popular Support Vector Machine classifier \cite{Zhang:2004}. 

Standard text classification distinguishes between labeled (training) documents and unlabeled (test) documents. Cross-language text classification poses an extra constraint in that training documents and test documents are written in different languages. Here, the language of the training documents is referred to as source language \sLang, and the language of the test documents is referred to as target language~\tLang. The vocabulary \Voc divides into \sVoc and \tVoc, called vocabulary of the source language and vocabulary of the target language, with $\sVoc\cap\tVoc=\emptyset$. I.e., documents from the training set and the test set map onto non-overlapping regions of the feature space. Thus, a linear classifier trained on \sData associates non-zero weights only with words from \sVoc, which in turn means that it cannot be used to classify documents written in~\tLang.

One way to overcome this ``feature barrier'' is to find a cross-lingual representation for documents written in \sLang and \tLang, which enables the transfer of classification knowledge between the two languages. 
Intuitively, one can understand such a cross-lingual representation as a concept space that underlies both languages. In the following, we will use \stMap to denote a map that associates the original $|V|$-dimensional representation of a document $d$ written in \sLang or \tLang with its cross-lingual representation.
Once such a mapping is found the cross-language text classification problem reduces to a standard classification problem in the cross-lingual space.
Note that the existing methods for cross-language text classification can be characterized by the way \stMap is constructed. For instance, cross-language latent semantic indexing \cite{Dumais:1997a} and cross-language explicit semantic analysis \cite{Potthast:2008} estimate \stMap using a parallel corpus. Other methods use linguistic resources such as a bilingual dictionary to obtain \stMap \cite{Bel:2003,Olsson:2005,Wu:2008}.

\section{Cross-Language Structural Correspondence Learning}
\label{sec:cross-language-structural-correspondence-learning}

\begin{figure}
\centering
\includegraphics{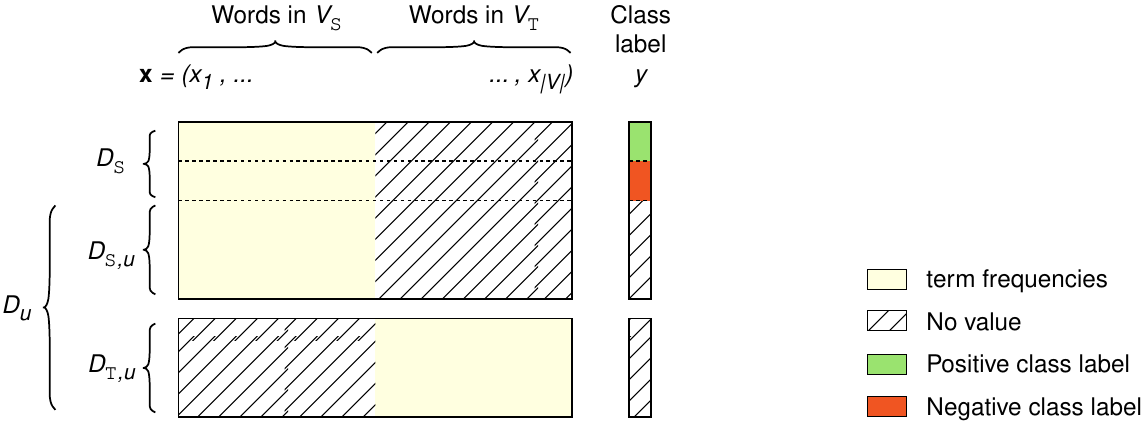} 
\caption{The document sets underlying CL-SCL. The subscripts $_\mathcal{S}$, $_\mathcal{T}$, and $_u$ designate ``source language'', ``target language'', and ``unlabeled''.} \label{cl-feature-space1}
\end{figure}

We now present a method for learning a map~\stMap by exploiting relations from unlabeled documents written in \sLang and \tLang. The proposed method, which we call cross-language structural correspondence learning, CL-SCL, addresses the following learning setup (see also Figure~\ref{cl-feature-space1}):

\begin{longenum}
\item
Given a set of labeled training documents $\sData$ written in language $\sLang$, the goal is to create a text classifier for documents written in a different language $\tLang$. We refer to this classification task as the \textit{target task}. An example for the target task is the determination of sentiment polarity, either positive or negative, of book reviews written in German (\tLang) given a set of training reviews written in English~(\sLang).
\item
In addition to the labeled training documents $\sData$ we have access to unlabeled documents \suData and \tuData from both languages $\sLang$ and~$\tLang$. Let \uData denote $\suData~\cup~\tuData$.
\item
Finally, we are given a budget of calls to a word translation oracle (e.g., a domain expert) to map words in the source vocabulary \sVoc to their corresponding translations in the target vocabulary \tVoc. For simplicity and without loss of applicability we assume here that the word translation oracle maps each word in \sVoc to exactly one word in \tVoc.
\end{longenum}

CL-SCL comprises three steps: In the first step, CL-SCL selects word pairs $\{\sWord,\tWord\}$, called pivots, where $\sWord\in\sVoc$ and $\tWord\in\tVoc$. Pivots have to satisfy the following conditions: 

\begin{describe}{{\em Confidence\quad}}
\item[{\em Confidence.}]
Both words, \sWord and \tWord, are predictive for the target task. 
\item[{\em Support.}]
Both words, \sWord and \tWord, occur frequently in \suData and \tuData, respectively. 
\end{describe}

The confidence condition ensures that, in the second step of CL-SCL, only those correlations are modeled that are useful for discriminative learning. The support condition, on the other hand, ensures that these correlations can be estimated accurately. Considering our sentiment classification example, the word pair $\{$excellent$_\sLang$, exzellent$_\tLang\}$ satisfies both conditions:
\Ni
the words are strong indicators of positive sentiment, and
\Nii
the words occur frequently in book reviews from both languages. Note that the support of \sWord and \tWord can be determined from the unlabeled data \uData. The confidence, however, can only be determined for \sWord since the setting gives us access to labeled data from \sLang only.

We use the following heuristic to form an ordered set \Pivots of pivots: First, we choose a subset $\pVoc$ from the source vocabulary \sVoc, $|\pVoc|\ll|\sVoc|$, which contains those words with the highest mutual information with respect to the class label of the target task in \sData. Second, for each word $\sWord\in\pVoc$ we find its translation in the target vocabulary \tVoc by querying the translation oracle; we refer to the resulting set of word pairs as the candidate pivots, $\Pivots'$\,:
\begin{equation*}
\Pivots' = \{\{\sWord,\Call{translate}{\sWord}\}\mid \sWord\in\pVoc\} .
\end{equation*}

We then enforce the support condition by eliminating in $\Pivots'$ all candidate pivots $\{\sWord, \tWord\}$ where the document frequency of \sWord in \suData or of \tWord in \tuData is smaller than some threshold $\phi$:
\begin{equation*}
\Pivots = \Call{candidateElimination}{\Pivots',\phi} .
\end{equation*}
Let $m$ denote $|\Pivots|$, the number of pivots.


In the second step, CL-SCL models the correlations between each pivot $\{\sWord,\tWord\}\in\Pivots$ and all other words $w\in\Voc\setminus \{\sWord,\tWord\}$. This is done by training linear classifiers that predict whether or not \sWord or \tWord occur in a document, based on the other words. For this purpose a training set $D_l$ is created for each pivot $p_l\in\Pivots$\,:
\begin{equation*}
D_l = \{(\Call{mask}{\mathbf{x}, p_l},\,\Call{in}{\mathbf{x}, p_l})\mid \mathbf{x} \in\uData\}
\end{equation*}

$\Call{mask}{\mathbf{x}, p_l}$ is a function that returns a copy of $\mathbf{x}$ where the components associated with the two words in $p_l$ are set to zero---which is equivalent to removing these words from the feature space. 
$\Call{in}{\mathbf{x}, p_l}$ returns +1 if one of the components of $\mathbf{x}$ associated with the words in $p_l$ is non-zero and -1 otherwise.
For each $D_l$ a linear classifier, characterized by the parameter vector $\mathbf{w}_l$, is trained by minimizing Equation~(\ref{eqn:regularized-training-error}) on $D_l$. Note that each training set $D_l$ contains documents from both languages. Thus, for a pivot $p_l = \{\sWord, \tWord\}$ the vector $\mathbf{w}_l$ captures both the correlation between \sWord and $\sVoc\setminus \{\sWord\}$ and the correlation between \tWord and $\tVoc\setminus \{\tWord\}$.

In the third step, CL-SCL identifies correlations {\em across} pivots by computing the singular value decomposition of the $|V| \times m$-dimensional parameter matrix $\mathbf{W}$, $\mathbf{W} = \begin{bmatrix} \mathbf{w}_1 & \dots & \mathbf{w}_m \end{bmatrix}$: 
\begin{equation*}
\label{eqn:svd}
\mathbf{U} \mathbf{\Sigma} \mathbf{V}^T \, = \ \text{SVD}(\mathbf{W}) .
\end{equation*}

Recall that $\mathbf{W}$ encodes the correlation structure between pivot and non-pivot words in the form of multiple linear classifiers. Thus, the columns of $\mathbf{U}$ identify common substructures among these classifiers. Choosing the columns of $\mathbf{U}$ associated with the largest singular values yields those substructures that capture most of the correlation in $\mathbf{W}$. We define \stMap as those columns of $\mathbf U$ that are associated with the $k$ largest singular values:
\begin{equation*}
\stMap = \mathbf{U}_{[1:k,\,1:|\Voc|]}^T .
\end{equation*}

Algorithm~\ref{alg:cl-scl} summarizes the three steps of CL-SCL. At training and test time, we apply the projection $\stMap$ to each input instance~$\mathbf{x}$. The vector $\mathbf{v}^*$ that minimizes the regularized training error for \sData in the projected space is defined as follows:
\begin{equation}
\label{eqn:regularized-training-error-projected}
\mathbf{v}^* = \argmin_{\mathbf{v}\in\real^k } \kern-3pt \sum_{(\mathbf{x},y) \in\sData} \kern-4pt \Loss(y,\,\mathbf{v}^T\stMap\mathbf{x}) + \lambda \Penalty(\mathbf{v}) .
\end{equation}

The resulting classifier, which will operate in the cross-lingual setting, is defined as follows:
\begin{equation*}
f(\mathbf{x}) = sign(\mathbf{v}^{*T} \stMap \mathbf{x}) .
\end{equation*}

\begin{algorithm}[t]
\caption{CL-SCL}\label{alg:cl-scl}

\renewcommand{\algorithmicindent}{2em}
\algsetblockx[Step]{Step}{EndStep}{3}{2em}
[3][Step]{\textbf{#2.} #3}

\begin{algorithmic}[0]

\smallskip
\Statex \parbox{8em}{\textbf{Input:}}		Labeled source data \sData
\Statex \parbox{8em}{\ }			Unlabeled data $\uData = \suData\cup\tuData$
\smallskip
\Statex \parbox{8em}{\textbf{Parameters:}}	$m$, $k$, $\lambda$, and $\phi$
\smallskip
\Statex \parbox{8em}{\textbf{Output:}}		$k\times |\Voc|$-dimensional matrix \stMap

\bigskip
\Step{1}{\Call{selectPivots}{\sData,\,$m$}}
\smallskip
\State $\pVoc = $ \Call{mutualInformation}{\sData}
\State $\Pivots' = \{\{\sWord,\Call{translate}{\sWord}\}\mid \sWord\in\pVoc\}$
\State $\Pivots = \Call{candidateElimination}{\Pivots',\phi}$
\EndStep

\Step{2}{\Call{trainPivotPredictors}{\uData,\,\Pivots}}
\smallskip
\For{$l= 1$ \textbf{to} $m$}
\State $D_l = \{(\Call{mask}{\mathbf{x}, p_l},\,\Call{in}{\mathbf{x}, p_l})\mid \mathbf{x} \in\uData\}$
\smallskip
\State $\mathbf{w}_l \! = \argmin_{\mathbf{w} \in \real^{|V|}} \limits \sum_{(\mathbf{x},y) \in D_l} \limits \! L(y,\mathbf{w}^T \mathbf{x})) + \lambda \Penalty(\mathbf{w})$
\EndFor
\State $\mathbf{W} = \begin{bmatrix} \mathbf{w}_1 & \dots & \mathbf{w}_m \end{bmatrix}$
\EndStep

\Step{3}{\Call{computeSVD}{$\mathbf{W}$,\,$k$}}
\smallskip
\State $\mathbf{U} \mathbf{\Sigma} \mathbf{V}^T = \text{SVD}(\mathbf{W})$
\State $\stMap = \mathbf{U}_{[1:k,\,1:|\Voc|]}^T $
\EndStep

\State \textbf{output} $\{\stMap\}$

\end{algorithmic}
\end{algorithm}

\subsection{An Alternative View of CL-SCL}

An alternative view of cross-language structural correspondence learning is provided by the framework of structural learning \cite{Ando:2005b}. The basic idea of structural learning is to constrain the hypothesis space, i.e., the space of possible weight vectors, of the target task by considering multiple different but related prediction tasks. In our context these auxiliary tasks are represented by the pivot predictors, i.e., the columns of $\mathbf{W}$. Each column vector $\mathbf{w}_l$ can be considered as a linear classifier which performs well in both languages. Thus, we can regard the column space of $\mathbf{W}$ as an approximation to the \textit{subspace of bilingual classifiers}. By computing $\text{SVD}(\mathbf{W})$ one obtains a compact representation of this column space in the form of an orthonormal basis $\stMap^T$. 

The subspace is used to constrain the learning of the target task by restricting the weight vector $\mathbf{w}$ to lie in the subspace defined by $\stMap^T$. Following \citeN{Ando:2005b} and \citeN{Quattoni:2007} we choose $\mathbf{w}$ for the target task to be $\mathbf{w}^* = \stMap^T \mathbf{v}^*$, where $\mathbf{v}^*$ is defined as follows:
\begin{equation}
\mathbf{v}^* = \argmin_{\mathbf{v}\in\real^k } \kern-3pt \sum_{(\mathbf{x},y) \in\sData} \kern-4pt \Loss(y, \,(\stMap^T \mathbf{v})^T \mathbf{x}) + \lambda \Penalty(\mathbf{v}).
\end{equation}

Since $(\stMap^T\mathbf{v})^T = \mathbf{v}^T\stMap$ it follows that this view of CL-SCL corresponds to the induction of a new feature space given by Equation~\ref{eqn:regularized-training-error-projected}.

\begin{figure}
\centering
\includegraphics[width=\columnwidth]{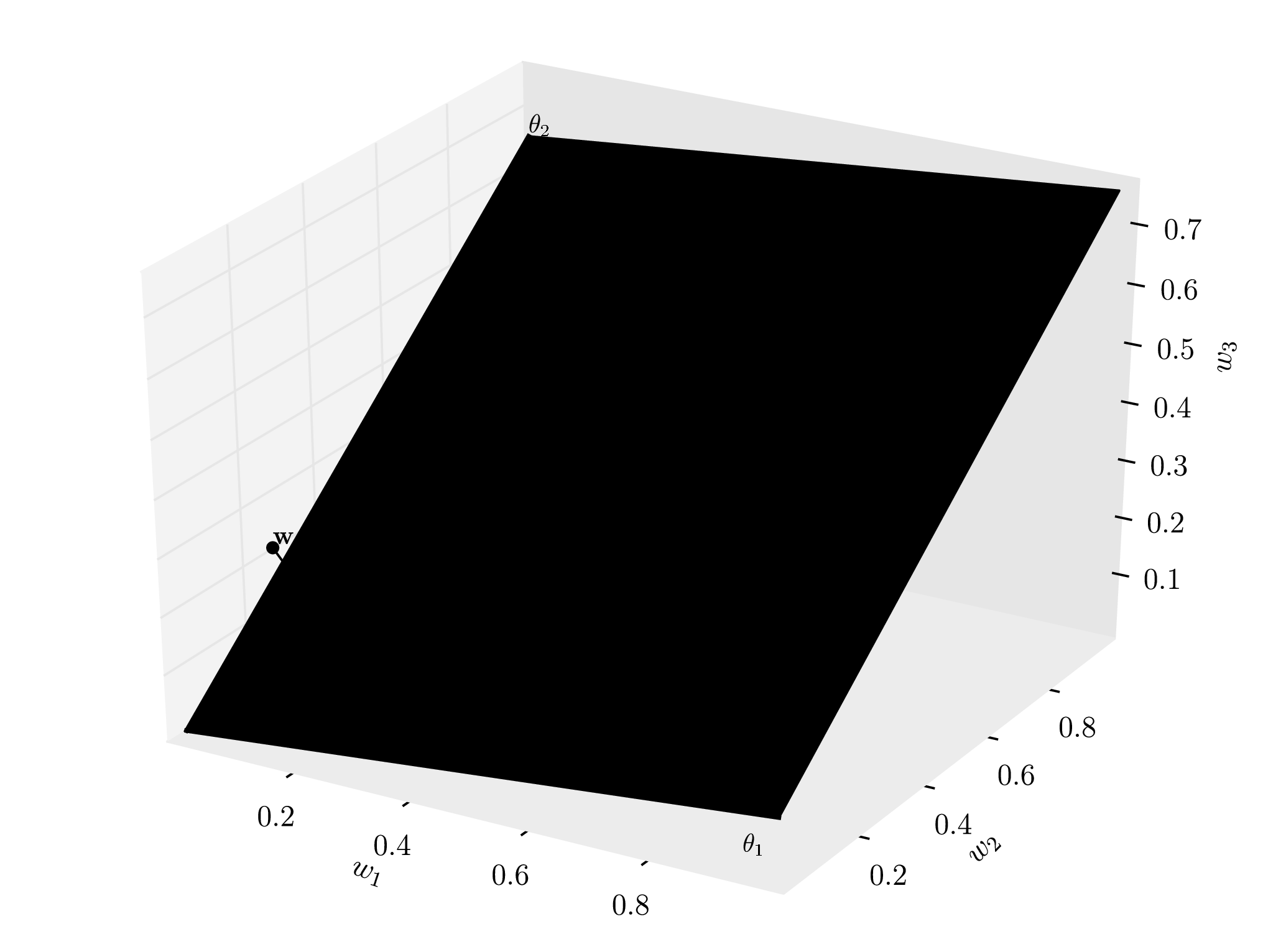}
\caption{Illustration of the subspace constraint for $|V|=3$ and $k=2$. The plane spanned by $\stMap_1$ and $\stMap_2$ shows the subspace induced by the two left singular vectors of $\mathbf{W} = [\mathbf{w}_1 \ \mathbf{w}_2 \ \mathbf{w}_3]$ associated with the largest singular values. For the target task, we restrict the weight vector $\mathbf{w}$ to lie in the subspace of the parameter space defined by $\stMap^T$, $\mathbf{w} = \stMap^T \mathbf{v}$.} 
\label{fig:subspace-constraint}
\end{figure}

Figure~\ref{fig:subspace-constraint} illustrates the basic idea of the subspace constraint for $|V|=3$ and $k=2$.


\subsection{Computational Considerations}
\label{sec:computational-considerations}

While the second step of CL-SCL involves the training of a fairly large number of linear classifiers, these classifiers can be learned very efficiently due to
\Ni
efficient learning algorithms for linear classifiers \cite{Shwartz:2007} and
\Nii
the fact that learning the pivot classifiers is an embarrassingly parallel problem. The computational bottleneck of the CL-SCL procedure is the SVD of the dense parameter matrix $\mathbf{W}$. In order to make the computation tractable, \citeN{Ando:2005b} as well as \citeN{Blitzer:2007} propose to set negative entries in $\mathbf{W}$ to zero, in order to obtain a sparse matrix for which the SVD can be computed more efficiently \cite{Berry:1992}. As a rational for this step the authors claim that the involved features ``yield much less specific information'' on the target concept, while ``positive weights are usually directly related to the target concept'' \cite{Ando:2005b}. 

We propose a different strategy to obtain a sparse parameter matrix $\mathbf{W}$, namely to enforce sparse pivot classifiers $\mathbf{w}_l$ by employing a proper regularization term \Penalty in the second step of CL-SCL. A straight-forward solution is to use L1 regularization \cite{Tibshirani:1996}, which imposes an L1-norm penalty on $\mathbf{w}$, $\Penalty(\mathbf{w}) = {\Vert \mathbf{w} \Vert}_1 = \sum_{j=1}^{|V|} |w_j|$. This strategy recently gained much attention in the natural language processing community; \citeN{Gao:2007} show that L1 regularized models have similar predictive power to L2 regularized models while being much smaller at the same time---i.e., less parameters are non-zero.

L1 regularization, however, has properties which are inadequate in the context of SCL, in particular its handling of highly correlated features. \citeN{Zou:2005} show that if there is a subset of features among which the pairwise correlations are high, L1 regularization tends to select only one feature while pushing the other feature weights to zero. This is certainly not desirable for SCL since it relies on the proper modeling of correlations in order to induce correspondences among features. L2 regularization, by contrast, exhibits such a grouping behavior, resulting in equal weights for correlated features. The Elastic Net combines both properties, the grouping behavior of L2 regularization and the sparsity property of L1 regularization \cite{Zou:2005}. It is given by the convex combination of both norms:

\begin{equation}
\label{eq:elastic-net}
\Penalty(\mathbf{w}) = \alpha {\Vert \mathbf{w} \Vert}_2^2 + (1-\alpha) {\Vert \mathbf{w} \Vert}_1 ,
\end{equation}
where $\alpha \in [0,1]$ models the trade-off between grouping and sparsity. The Elastic Net is widely used in bioinformatics, in particular the study of gene expression, however, its use for applications in natural language processing or information retrieval has not been studied yet. 


\section{Experiments}
\label{sec:experiments}

We evaluate CL-SCL for the task of cross-language sentiment and topic classification using English as source language and German, French, and Japanese as target languages. We first describe the experimental design and give implementation details, we then present the evaluation results and, finally, we report on detailed analyses with respect to the nature of the induced cross-lingual correspondences, the use of unlabeled data, and important hyperparameters including the impact of different regularization methods.

\subsection{Datasets}

We use the cross-lingual sentiment dataset provided by \citeN{Prettenhofer:2010}.%
\footnote{Available at \url{http://www.webis.de/research/corpora/webis-cls-10/}}
The dataset contains Amazon product reviews for the three product categories books, dvds, and music in the languages English, German, French and Japanese. Each document is labeled according to its sentiment polarity as either positive or negative. The documents in the dataset are organized by language and product category. For each language-category pair there are three balanced disjoint sets of training, test, and unlabeled documents; the respective set sizes are 2,000, 2,000, and 9,000-50,000. Similar to \citeN{Prettenhofer:2010}, each document $d$ is represented as a normalized (unit length) feature vector $\mathbf{x}$ under a unigram bag-of-words model. Based on this dataset we create two tasks (see Table~\ref{tab:table-dataset-statistics} for a summary statistics):

\paragraph*{Sentiment Classification Task}

For the task of cross-language sentiment classification the original partitioning of the cross-lingual sentiment dataset is used. Analogous to \citeN{Prettenhofer:2010} English is employed as source language, and German, French, and Japanese as target languages. For each of the nine target-language-category-combinations a sentiment classification task is created by taking the training set and the unlabeled set for some product category from \sLang and the test set and the unlabeled set for the same product category from \tLang.

\paragraph*{Topic Classification Task (Product Categories)}

For the task of cross-language topic classification we discard the original sentiment labels and use the product category, i.e., books, dvd, and music as the document label. Again we use English as the source language and German, French, and Japanese as target languages. Note that---in contrast to the sentiment classification tasks---classifying reviews according to product categories is a multi-class classification problem with three mutually exclusive classes. Hence for each of the three target languages a cross-language topic classification task is created by combining the training set and the unlabeled set of each product category from \sLang with the test set and the unlabeled set of each product category from \tLang. For each of the three tasks we have 6,000 training and 6,000 test documents, each containing a balanced number of examples.

\begin{acmtable}{0.9\columnwidth}
\centering

\begin{tabular*}{0.9\columnwidth}{@{}ll  r@{\ \ }r  r@{\ \ }r r@{\ \ }r}
 \raisebox{-2ex}{\tLang} & \raisebox{-2ex}{\bfseries Category} & \multicolumn{2}{c}{\bfseries Unlabeled data} & \multicolumn{2}{c}{\bfseries Labeled data} & \multicolumn{2}{c}{\bfseries Vocabulary}  \\[-2ex]
 \cmidrule(lr){3-4}  \cmidrule(lr){5-6} \cmidrule(lr){7-8} 
\addlinespace[-0.5ex]
  & & $|\suData|$ & $|\tuData|$ & $|\sData|$ & \multicolumn{1}{c}{$|\tData|$} & $|\sVoc|$ & \multicolumn{1}{c}{$|\tVoc|$} \\ 
\hline
\addlinespace[2pt]
&books & 50,000 & 50,000 & 2,000 & 2,000 & 64,682 & 108,573\\
German&dvd & 30,000 & 50,000&  2,000 & 2,000 & 52,822 & 103,862\\
&music& 25,000 & 50,000 &  2,000 & 2,000 & 41,306 & 99,287\\
\addlinespace
&books& 50,000 & 32,000 &  2,000 & 2,000 & 64,682 & 55,016\\
French&dvd& 30,000 & \ \ 9,000 &  2,000 & 2,000 & 52,822 & 29,519\\
&music& 25,000 & 16,000 &  2,000 & 2,000 & 41,306 & 42,097\\
\addlinespace
&books& 50,000 & 50,000 &  2,000 & 2,000 & 64,682 & 52,311\\
Japanese& dvd & 30,000 & 50,000 &  2,000 & 2,000 & 52,822 & 54,533\\
&music& 25,000 & 50,000 &  2,000 & 2,000 & 41,306 & 54,463\\
\addlinespace[2pt]
\hdashline[2pt/2pt]
\addlinespace[2pt]
German&- & 60,000 & 60,000&  6,000 & 6,000 & 76,629 & 124,529\\
French&-& 60,000 & \ \ 45,000 &  6,000 & 6,000 & 76,629 & 74,807\\
Japanese& - & 60,000 & 60,000 &  6,000 & 6,000 & 76,629 & 64,050\\
\hline
\addlinespace[2pt]
\end{tabular*}

\raggedright
Summary statistics for the nine cross-language sentiment classification tasks (first nine rows) and the three cross-language topic classification tasks (last three rows). $|\suData|$ and $|\tuData|$ give the number of unlabeled documents from \sLang and \tLang; $|\sData|$ and $|\tData|$ give the number of training and test documents. All document sets are balanced. 
\caption{Dataset statistics.}
\label{tab:table-dataset-statistics} 
\end{acmtable}

\subsection{Implementation}

Within all experiments we employ linear classifiers, which are trained by minimizing Equation~(\ref{eqn:regularized-training-error}) using a stochastic gradient descent (SGD) algorithm. In particular, we use the plain SGD algorithm as described by \citeN{Zhang:2004} while adopting the learning rate schedule from PEGASOS \cite{Shwartz:2007}. Analogous to \citeN{Blitzer:2007} and \citeN{Ando:2005b} we employ as loss function \Loss the modified Huber loss \cite{Zhang:2004}, a smoothed version of the hinge loss:
\begin{equation}
\Loss(y,p)=\begin{cases}
\text{max}(0,1-py)^2, & \text{if }py \ge -1 \\
-4py, & \text{otherwise.}
\end{cases}
\end{equation} 

SGD and related methods based on stochastic approximation have been successfully applied to solve large-scale linear prediction problems in natural language processing and information retrieval \cite{Zhang:2004,Shwartz:2007}. Their major advantages are efficiency and ease of implementation. 

SGD, however, cannot be applied directly in connection with L1 regularization (and thus the Elastic Net) due to the fluctuations of the approximated gradients. To overcome this problem different solutions have been proposed, in particular methods based on truncated gradients \cite{Langford:2009,Tsuruoka:2009} and projected gradients \cite{Duchi:2008}. In our experiments we employ the truncated stochastic gradient algorithm proposed by \citeN{Tsuruoka:2009}, which uses the cumulative L1 penalty to smooth out fluctuations in the approximated gradients.%
\footnote{Our implementation is available at \url{http://github.com/pprett/bolt}}
Note that Elastic Net regularization is applied for the pivot classifiers only, all other classifiers are trained using L2 regularization. 

SGD receives two hyperparameters as input: the number of iterations $T$, and the regularization parameter $\lambda$. In our experiments $T$ is always set to $10^6$, which is about the number of iterations required for SGD to converge. For the target task, $\lambda$ is determined by 3-fold cross-validation, testing for $\lambda$ all values $10^{-i}, i\in [0;6]$. For the pivot prediction task, $\lambda$ is set to the small value of $10^{-5}$, in order to favor model accuracy over generalizability.

Since SGD is sensitive to feature scaling the projection $\stMap \mathbf{x}$ is post-processed as follows:
\Ni
Each feature of the cross-lingual representation is standardized to zero mean and unit variance, where mean and variance are estimated on $\sData\cup\uData$.
\Nii
The cross-lingual document representations are scaled by a constant $\alpha$ such that ${\vert\sData\vert}^{-1} \sum_{\mathbf{x}\in\sData}\|\alpha\stMap\mathbf{x}\| = 1$.

For multi-class classification the one-against-all-strategy is applied. For multi-class problems, the set of pivot candidates \pVoc is formed as follows:
\Ni
rank for each class the words according to mutual information with respect to all other classes, and
\Nii
select from each ranking those words with the highest mutual information. 

We use the bilingual dictionary provided by \citeN{Prettenhofer:2010} as word translation oracle. If the source word is not contained in the dictionary we resort Google Translate, which returns a single translation for each query word.%
\footnote{\url{http://translate.google.com}}
Note that the word translation oracle operates context-free, which is suboptimal; however, we do not sanitize the translations to demonstrate the robustness of CL-SCL with respect to translation noise.

\subsection{Upper Bound and Baseline}

To get an upper bound on the performance of a cross-language method we first consider the monolingual setting. For each task a linear classifier is learned on the training set of the target language and tested on the test set. The resulting accuracy scores are referred to as upper bound; this bound informs us about the expected performance on the target task if training data in the target language is available. 

We choose a machine translation baseline to compare CL-SCL to another cross-language method. Statistical machine translation technology offers a straightforward solution to the problem of cross-language text classification and has been used in a number of cross-language sentiment classification studies \cite{Hiroshi:2004,Bautin:2008,Wan:2009}. Our baseline CL-MT is determined as follows:
\Ni
learn a linear classifier on the training data, and
\Nii
translate the test documents into the source language,
\Niii
predict the sentiment polarity of the translated test documents. 

Translations of the test documents into the source language via Google Translate are provided by \citeN{Prettenhofer:2010}. Note that the baseline CL-MT does not make use of unlabeled documents.

\subsection{Experimental Results}

Table~\ref{tab:table-cls-performance} contrasts the classification performance of CL-SCL with the upper bound and the baseline. Due to the inherent randomness of the training algorithm, we report the accuracy scores as mean $\mu$ and standard deviation $\sigma$ of ten repetitions of SGD. We use McNemar's test to analyze whether or not the results of CL-SCL and CL-MT are statistically significant \cite{Dietterich:1998}. Again, due to the randomness of the training algorithm statistical significance is analyzed for each of the ten repetitions, whereas significance at a specific level is reported only if it applies to all repetitions.

Observe that the upper bound does not exhibit high variability across the three languages. For sentiment classification the average accuracy is about 82\%, which is consistent with prior work on monolingual sentiment analysis \cite{Pang:2002,Blitzer:2007}. For product category classification the average accuracy is in the low 90's, which is also consistent with prior work on monolingual product category classification \cite{Crammer:2009}. 

The performance of CL-MT, however, differs considerably between the two European languages and Japanese: for Japanese, the averaged differences between the upper bound and CL-MT (9.5\%, 7.3\%) are much larger than for German and French (5.3\%, 1.7\%). This can be explained by the fact that machine translation works better for European than for Asian languages such as Japanese.

Recall that CL-SCL receives four hyperparameters as input: the number of pivots $m$, the dimensionality of the cross-lingual representation $k$, the minimum support $\phi$ of a pivot in \suData and \tuData, and the Elastic Net coefficient $\alpha$. For cross-language sentiment classification we use fixed values of $m = 450$, $k = 100$, $\phi = 30$, and $\alpha = 0.85$. For cross-language topic classification we found that smaller values of $m$ and $k$ work significantly better. The results for topic classification are obtained by using fixed values of $m = 250$, $k = 50$, $\phi = 50$, and $\alpha = 0.85$. The parameter settings have been optimized using the German book review task (sentiment) and the German task (topic).

The results show that CL-SCL either outperforms CL-MT or is at least competitive across all tasks. For German and Japanese sentiment classification we observe significant differences at a 0.05 and 0.001 confidence level. For product category classification we observe significant differences only for Japanese (0.001 confidence level). Interestingly, for German music reviews, the accuracy of CL-SCL even surpasses the upper bound; this can be interpreted as a semi-supervised learning effect that stems from the massive use of unlabeled data. The rightmost column of Table~\ref{tab:table-cls-performance} shows the relative reduction in error due to cross-lingual adaptation of CL-SCL over CL-MT. CL-SCL reduces the relative error by an average of 59\% (sentiment classification) and 30\% (topic classification) over CL-MT.

\begin{acmtable}{\columnwidth}
\centering
\newcommand{\sig}{\ensuremath{\dagger}\xspace}
\newcommand{\dsig}{\ensuremath{\dagger\dagger}\xspace}

\begin{tabular*}{\columnwidth}{@{}ll  cc  *{2}{*{3}{r@{\ \ }r@{\ \ }r}}r@{}r@{}}
\raisebox{-2ex}{\tLang} & \raisebox{-2ex}{\bfseries Cat.} &  \multicolumn{2}{c}{\bfseries Upper Bound} & \multicolumn{3}{c}{\bfseries CL-MT} & \multicolumn{3}{c}{\bfseries CL-SCL}  \\[-2ex]
\cmidrule(lr){3-4} \cmidrule(lr){5-7} \cmidrule(lr){8-10}
\addlinespace[-0.5ex]
 & & $\mu$ & $\sigma$ & \multicolumn{1}{c}{$\mu$} & \multicolumn{1}{c}{$\sigma$} & \multicolumn{1}{c}{$\Delta$} & \multicolumn{1}{c}{$\mu$} & \multicolumn{1}{c}{$\sigma$} & \multicolumn{1}{c}{$\Delta$} & RR[\%]\\ 
\hline
\addlinespace[2pt]
&books &83.79&$\pm$0.20&79.68&$\pm$0.13&4.11&\ \sig\bf83.34&$\pm$0.02&0.45&89.05\\
German&dvd &81.78&$\pm$0.27&77.92&$\pm$0.25&3.86&\ \sig\bf80.89&$\pm$0.02&0.89&76.94\\
&music&82.80&$\pm$0.13&77.22&$\pm$0.23&5.58&\ \sig\bf82.90&$\pm$0.00&-0.10&101.79\\
&books&83.92&$\pm$0.14&80.76&$\pm$0.34&3.16&\ \ \bf81.27&$\pm$0.08&2.65&16.14\\
French&dvd&83.40&$\pm$0.28&78.83&$\pm$0.19&4.57&\ \ \bf80.43&$\pm$0.05&2.97&35.01\\
&music&86.09&$\pm$0.13&75.78&$\pm$0.65&10.31&\ \ \bf78.05&$\pm$0.06&8.04&22.02\\
&books&78.09&$\pm$0.14&70.22&$\pm$0.27&7.87&\dsig\bf77.00&$\pm$0.06&1.09&86.15\\
Japanese& dvd &81.56&$\pm$0.28&71.30&$\pm$0.28&10.26&\dsig\bf76.37&$\pm$0.05&5.19&49.42\\
&music&82.33&$\pm$0.13&72.02&$\pm$0.29&10.31&\dsig\bf77.34&$\pm$0.06&4.99&51.60\\
\addlinespace[2pt]
\hdashline[2pt/2pt]
\addlinespace[2pt]
German&-&92.95&$\pm$0.11&92.25&$\pm$0.07&0.70&\ \ \bf92.61&$\pm$0.06&0.34&51.43\\
French&-&93.27&$\pm$0.07& \bf90.58&$\pm$0.17&2.69&\ \ 90.57&$\pm$0.13&2.70&-0.37\\
Japanese&-&89.43&$\pm$0.11&82.14&$\pm$0.22&7.29&\dsig\bf85.03&$\pm$0.10&4.40&39.64\\
\hline
\addlinespace[2pt]
\end{tabular*}

\raggedright
Evaluation results for sentiment classification (first nine rows) and topic classification (last three rows). Accuracy scores (mean $\mu$ and standard deviation $\sigma$ of 10 repetitions of SGD) on the test set of the target language \tLang are reported. $\Delta$ gives the difference in accuracy to the upper bound. Statistical significance (McNemar) of CL-SCL is measured against CL-MT (\sig indicates 0.05 and \dsig 0.001). RR gives the relative reduction in error over CL-MT. For sentiment classification, CL-SCL uses $m=450$, $k=100$, $\phi=30$, and $\alpha=0.85$. 
For topic classification, CL-SCL uses $m=250$, $k=50$, $\phi=50$, and $\alpha=0.85$. 
\caption{Cross-language sentiment and topic classification results. }
\label{tab:table-cls-performance}
\end{acmtable}

\subsection{Sensitivity Analysis}

CL-SCL receives a number of hyperparameters as input; the purpose of this section is to elaborate on each hyperparameter. In the following, we will analyze the sensitivity of each hyperparameter in isolation while keeping the others fixed. If not specified otherwise, we use the same setting of the hyperparameters as in Table~\ref{tab:table-cls-performance}.

\begin{figure*}[t]
\includegraphics[width=0.32\textwidth]{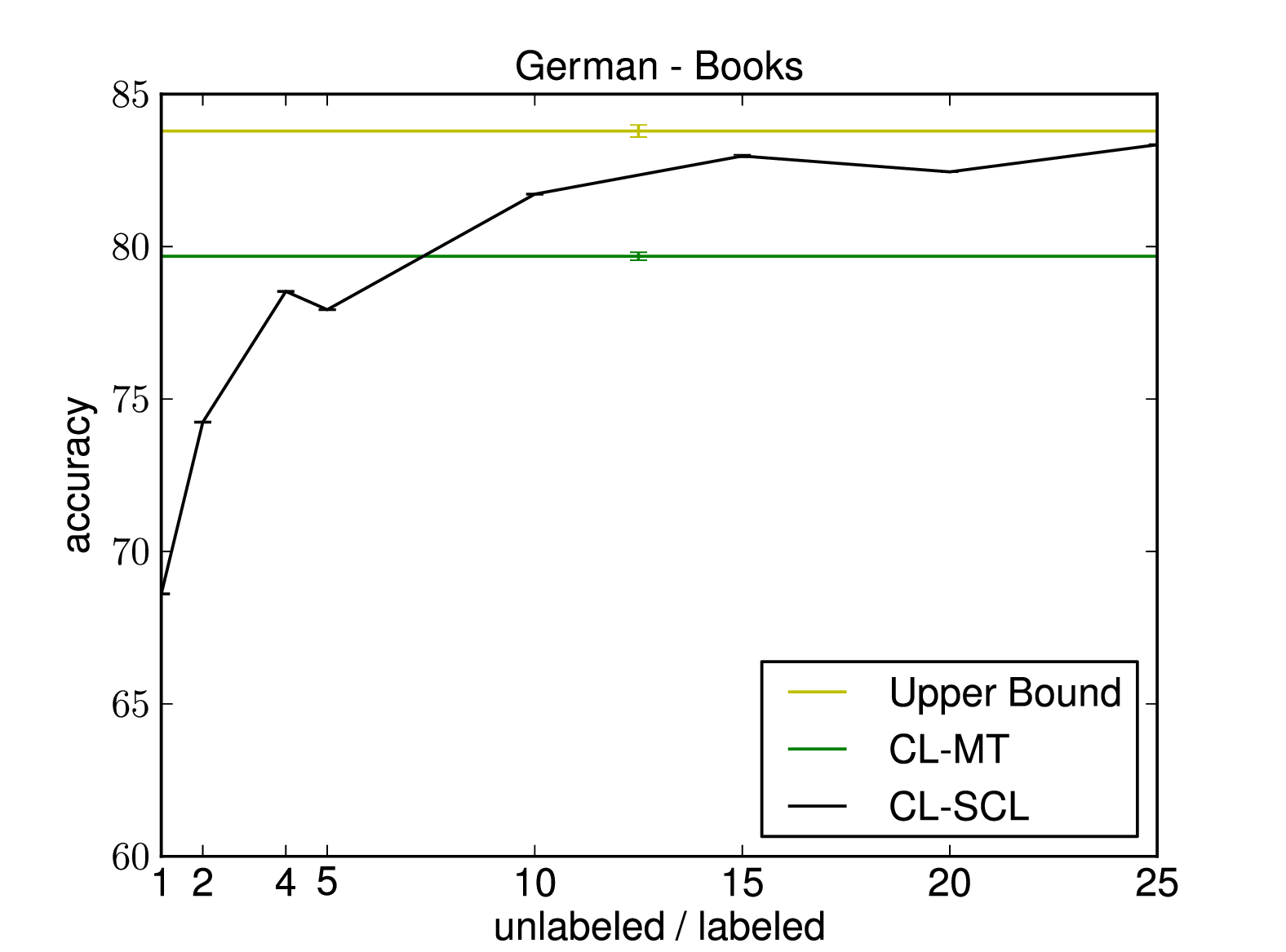}
\includegraphics[width=0.32\textwidth]{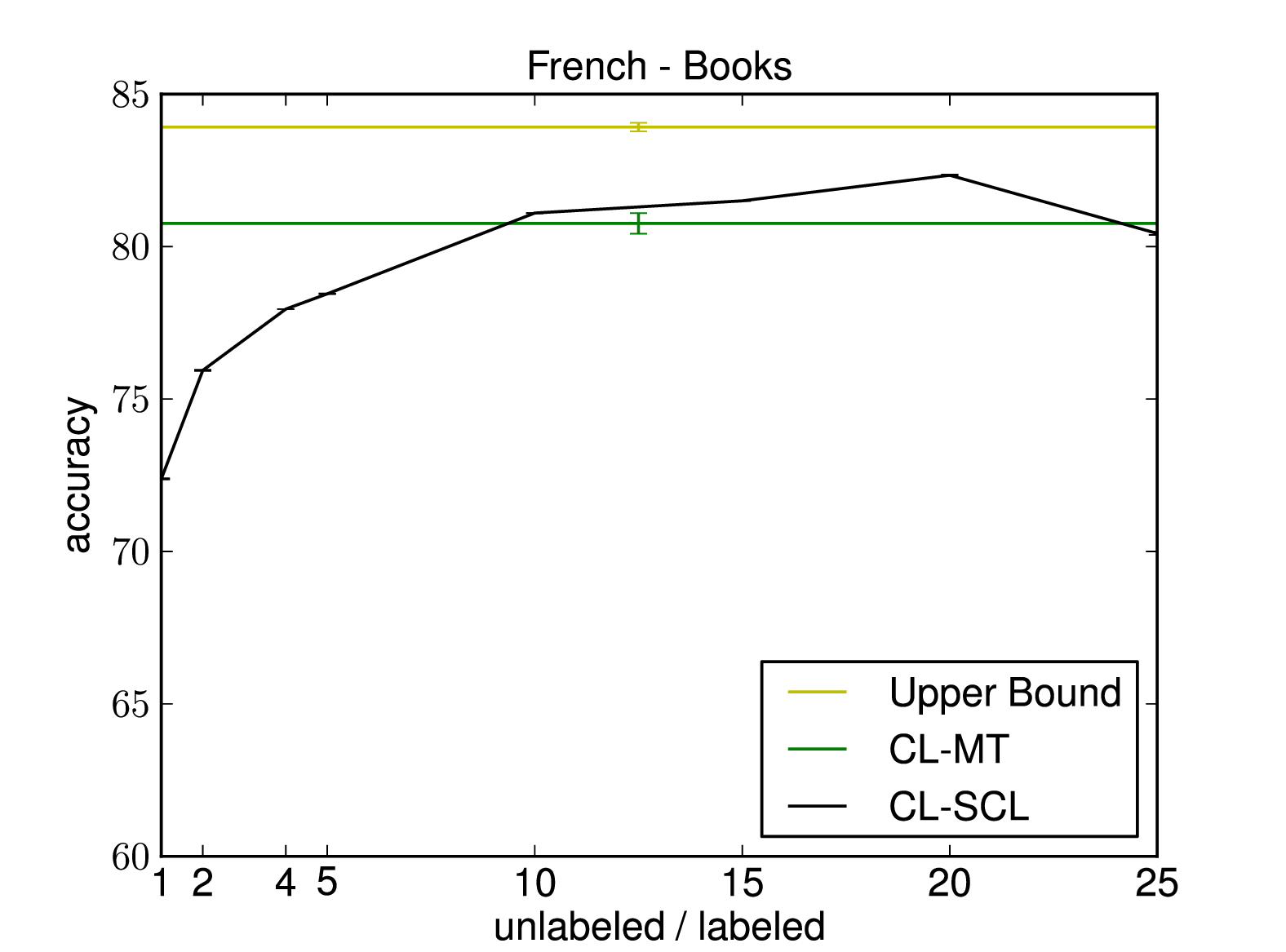}
\includegraphics[width=0.32\textwidth]{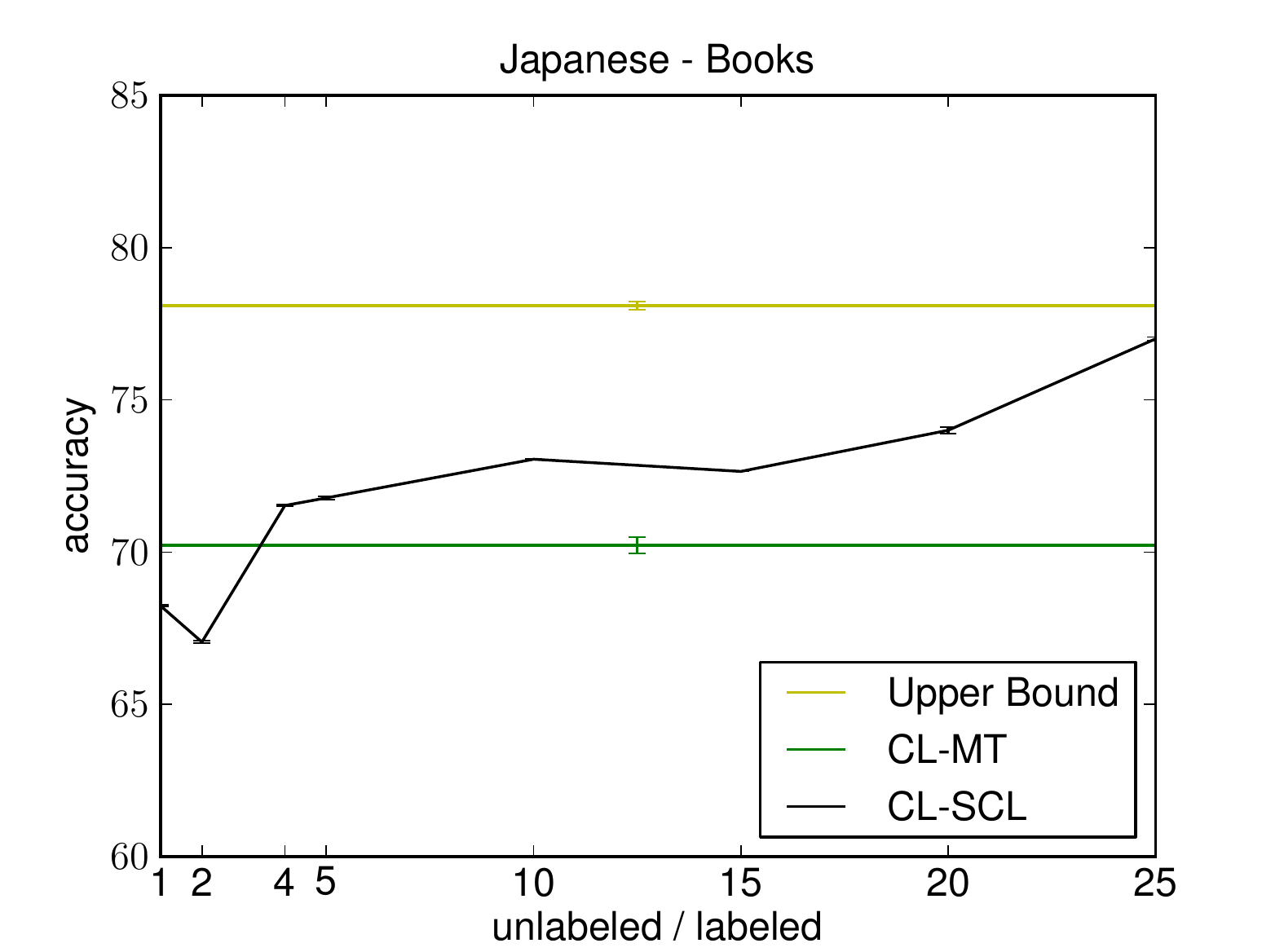} \\[2ex]
\includegraphics[width=0.32\textwidth]{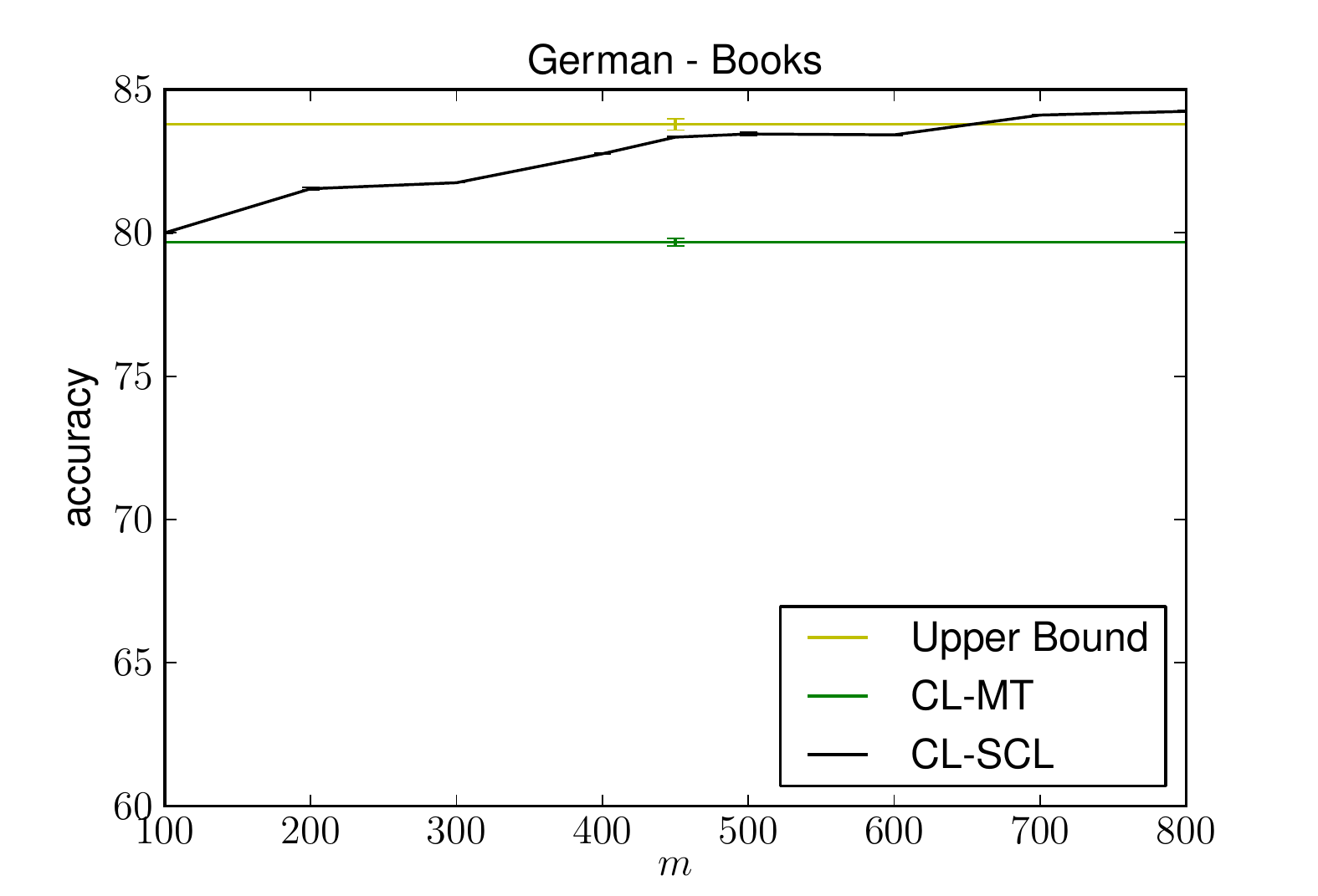}
\includegraphics[width=0.32\textwidth]{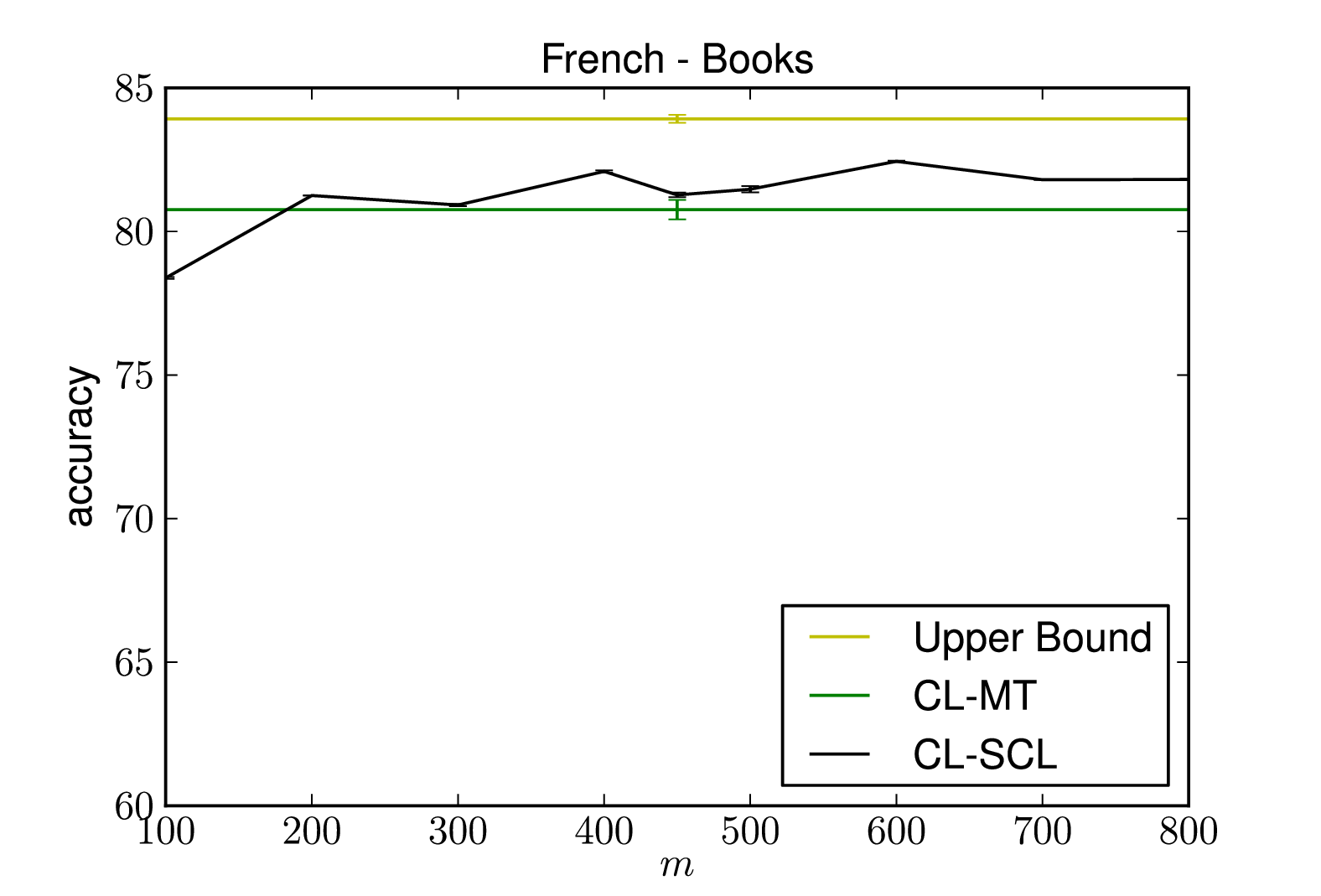}
\includegraphics[width=0.32\textwidth]{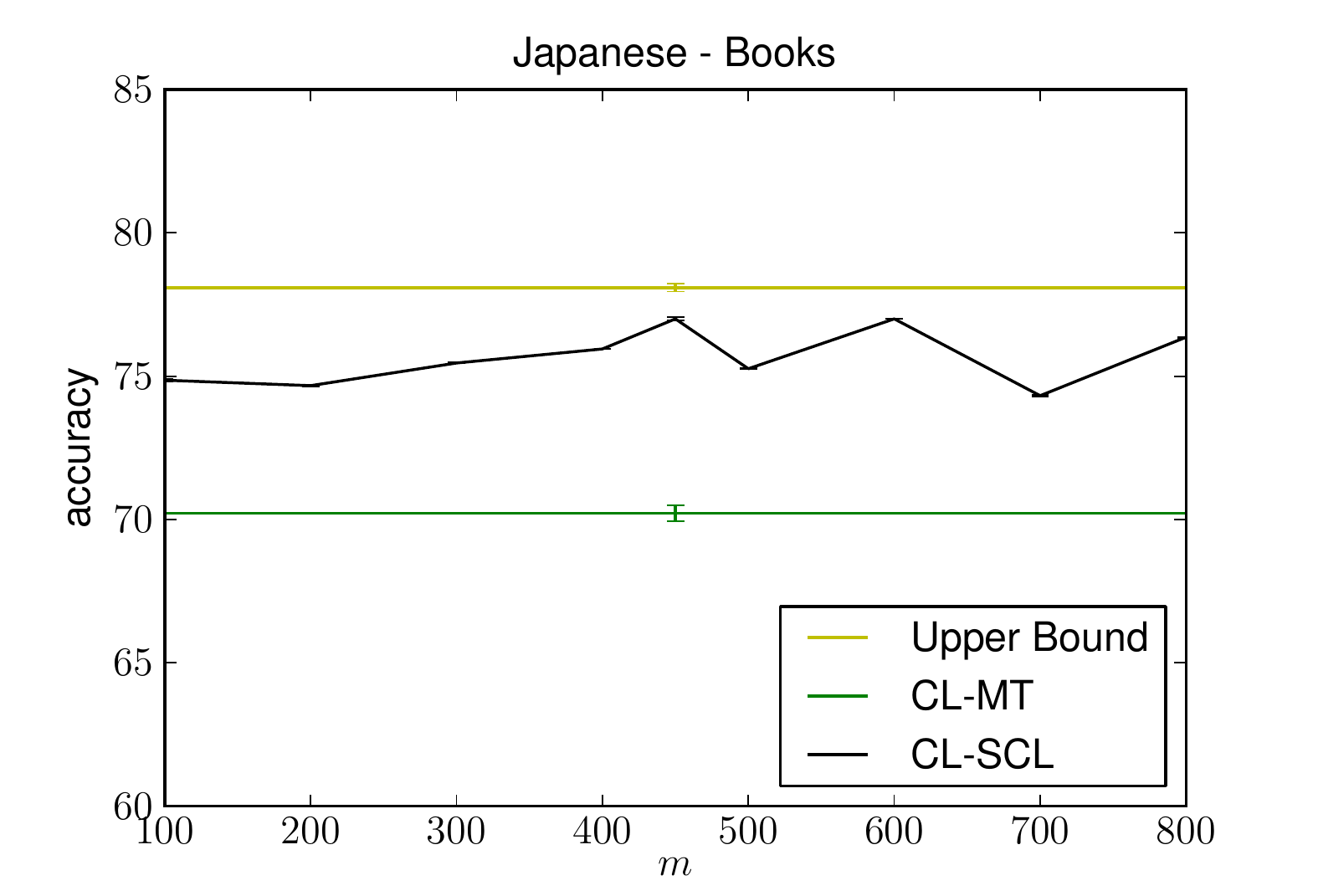} \\[2ex]
\includegraphics[width=0.32\textwidth]{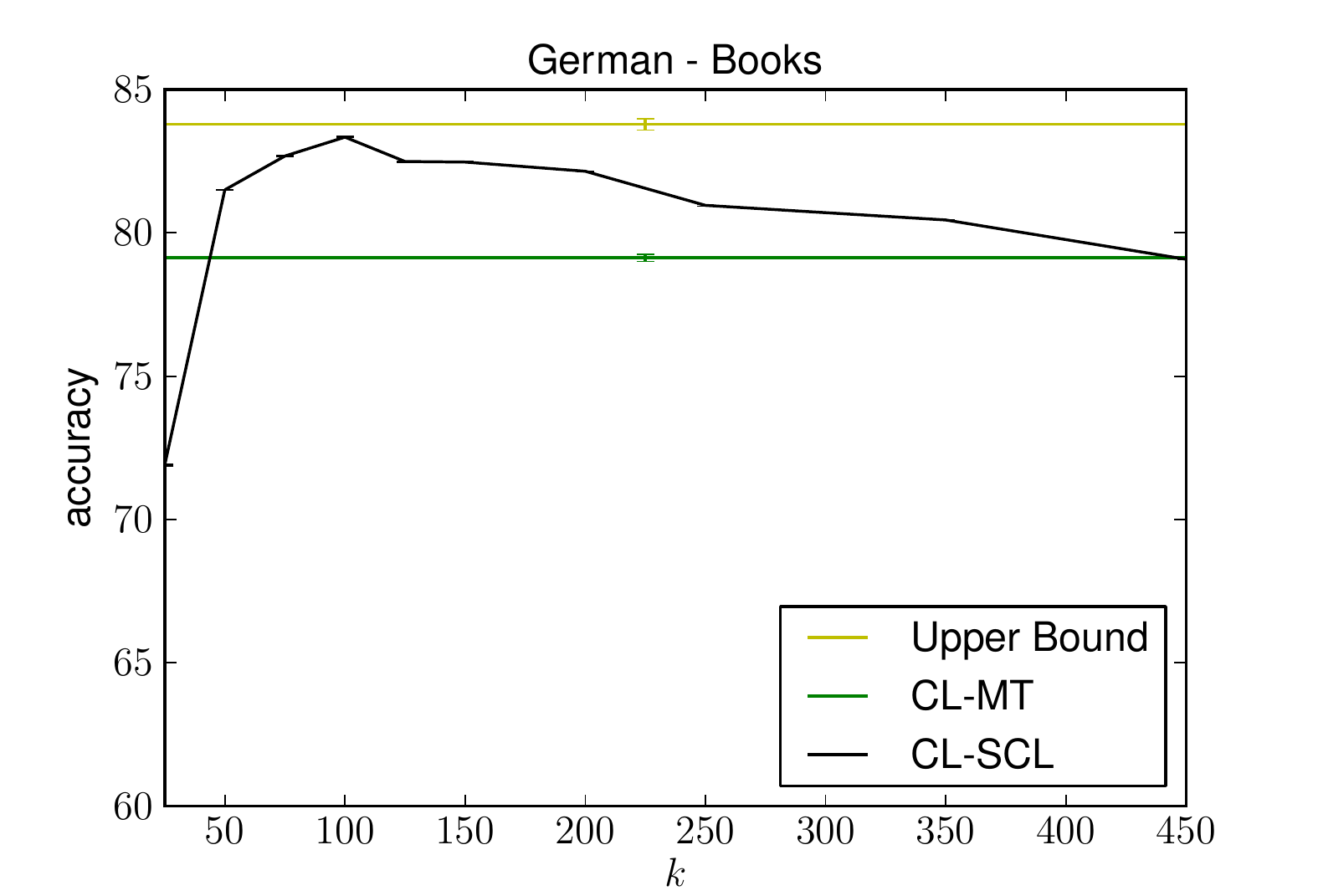}
\includegraphics[width=0.32\textwidth]{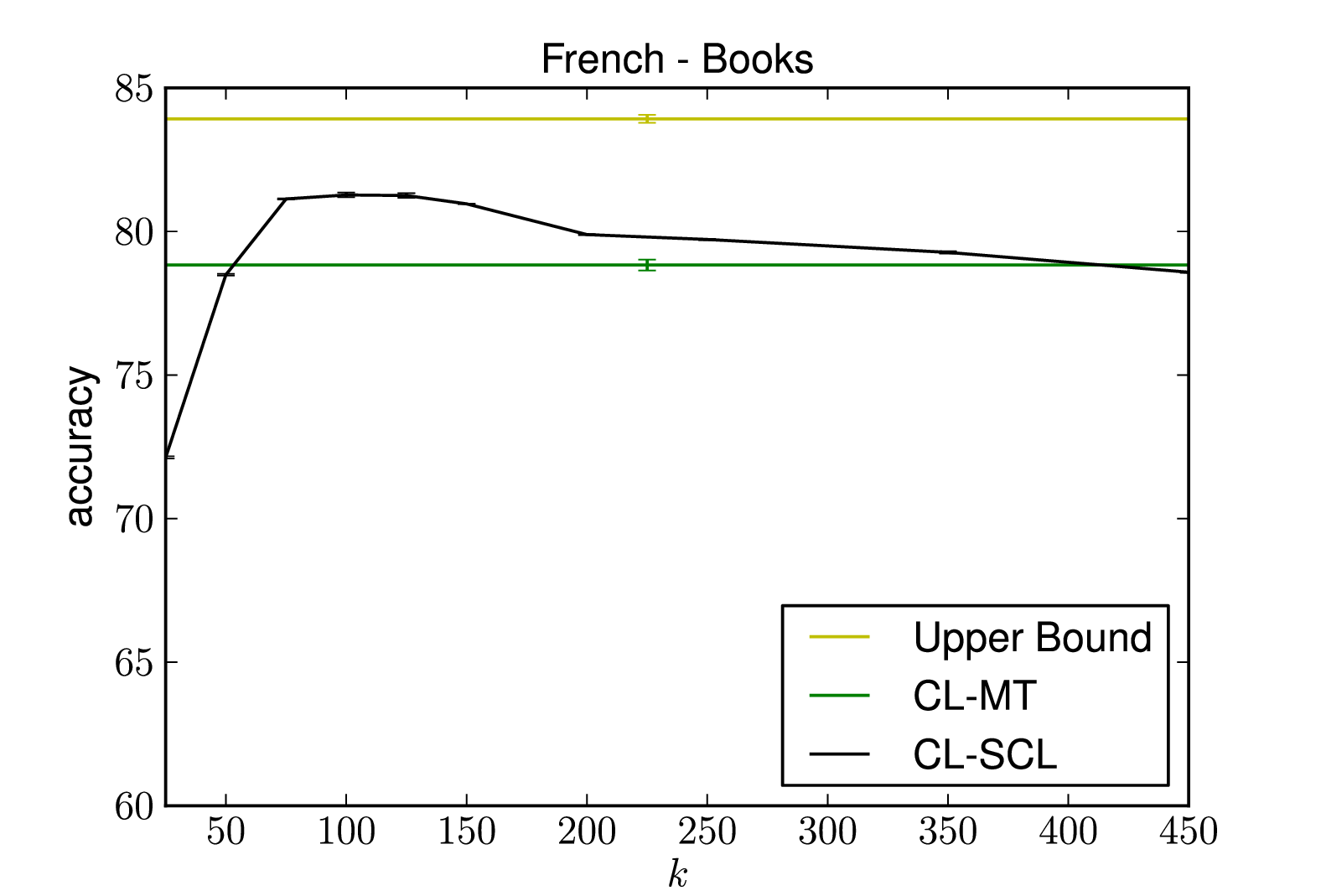}
\includegraphics[width=0.32\textwidth]{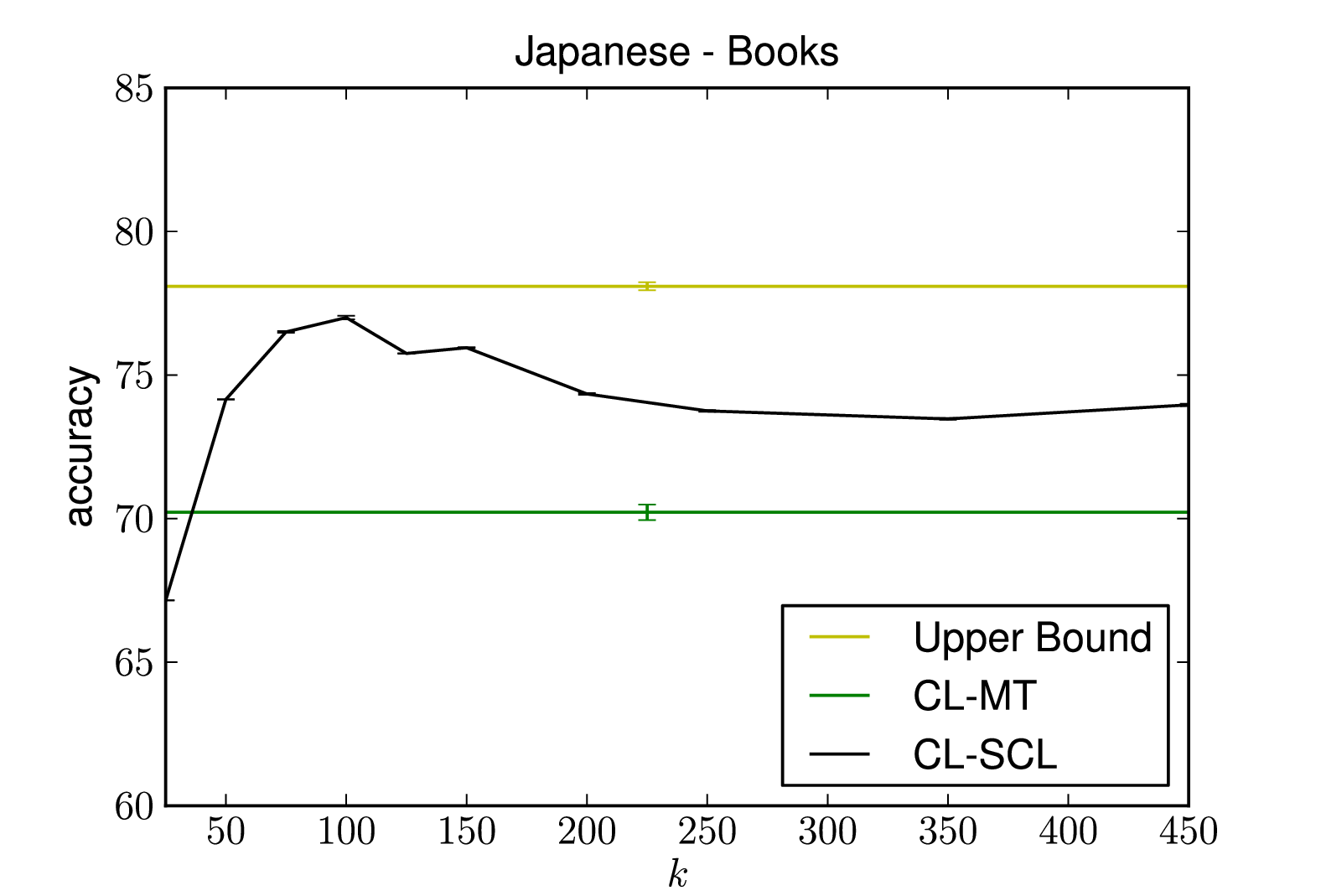}

\caption{Influence of unlabeled data and hyperparameters on the performance of CL-SCL. The rows show the performance of CL-SCL as a function of \Ni the ratio between labeled and unlabeled documents, \Nii the number of pivots $m$, and \Niii the dimensionality of the cross-lingual representation $k$.} \label{fig:parameter-influence}
\end{figure*}

\paragraph*{Unlabeled Data}

The first row of Figure~\ref{fig:parameter-influence} shows the performance of CL-SCL as a function of the ratio of labeled and unlabeled documents for sentiment classification of book reviews. A ratio of~1 means that $|\suData| = |\tuData| = 2{,}000$, while a ratio of $25$ corresponds to the setting of Table~\ref{tab:table-cls-performance}. As expected, an increase in the number of unlabeled documents results in an improved performance. However, a saturation at a ratio of 10 can be observed across most tasks.

\paragraph*{Number of Pivots}

The second row shows the influence of the number of pivots $m$ on the performance of CL-SCL. Compared to the size of the vocabularies \sVoc and \tVoc, which is in $10^5$ order of magnitude, the number of pivots is very small. The plots show that even a small number of pivots captures a significant amount of the correspondence between \sLang and \tLang.

\paragraph*{Dimensionality of the Cross-Lingual Representation}

The third row shows the influence of the dimensionality of the cross-lingual representation $k$ on the performance of CL-SCL. Obviously the SVD is crucial to the success of CL-SCL if $m$ is sufficiently large. Observe that the value of $k$ is task-insensitive: a value of $50$$<$$k$$<$$150$ works equally well across all tasks.

\paragraph*{Effect of Regularization}

Table~\ref{tab:table-regularization} compares the effect of three different regularization terms on the performance of CL-SCL. The third column, $\text{L2}^+$, refers to the strategy in \cite{Blitzer:2006} and \cite{Prettenhofer:2010} with ordinary L2 regularization and negative weights set to zero. The fifth column shows the performance of L1 regularization. Observe that L1 regularization drastically reduces the number of non-zero features, from 16\% to 2\% on average. We argued in Section~\ref{sec:computational-considerations} that L1 regularization is not adequate due to its improper handling of highly correlated features and we proposed the Elastic Net penalty as an alternative. The empirical evidence supports this claim: Elastic Net regularization consistently outperforms both $\text{L2}^+$ and L1 regularization while keeping the number of non-zero features low (15\% on average). Note that Elastic Net regularization adds an additional hyperparameter $\alpha$ that trades off the relative importance of L2 and L1 regularization. In the above experiments the value of $\alpha$ is chosen such that the obtained density roughly equals the density of $\text{L2}^+$. A convenient property of the Elastic Net is that it encompasses L2 and L1 regularization as special cases (either $\alpha=1$ or $\alpha=0$). Thus, if $m$ and $|V|$ are sufficiently small and a dense SVD is computationally feasible $\alpha=1$ is optimal. Otherwise, the optimal choice of $\alpha$ is governed by the computing resource. 

The use of Elastic Net regularization to obtain sparse pivot classifiers has implications beyond CL-SCL, in particular for the application of Alternating Structural Optimization \cite{Ando:2005a} and Structural Correspondence Learning \cite{Blitzer:2006} in high dimensional feature spaces.

\begin{acmtable}{\columnwidth}

\centering
\renewcommand{\tabcolsep}{17pt}
\newcommand{\sig}{\ensuremath{\dagger}\xspace}
\newcommand{\dsig}{\ensuremath{\dagger\dagger}\xspace}

\begin{tabular*}{\columnwidth}{@{}cl  *{3}{*{2}{c@{\ \ }r}}}
\raisebox{-2ex}{\tLang} & \kern-0.5em\raisebox{-2ex}{\bfseries Category} &  \multicolumn{2}{c}{\bfseries $\text{L2}^+$} & \multicolumn{2}{c}{\bfseries L1} & \multicolumn{2}{c}{\bfseries Elastic Net}  \\[-2ex]
\cmidrule(lr){3-4} \cmidrule(lr){5-6} \cmidrule(lr){7-8}
\addlinespace[-0.5ex]
 & &  $\mu$ & d[\%] & $\mu$ &  d[\%] & $\mu$ & d[\%]\\ 
\hline
\addlinespace[2pt]
&books &79.50&17.88&82.45&1.24& \bf83.34&11.02\\
German&dvd &77.06&16.84&78.60&1.43& \bf80.89&12.25\\
&music&77.60&16.00&81.41&1.72& \bf82.90&13.92\\
\addlinespace
&books&79.02&16.50&80.75&1.87& \bf81.27&14.13\\
French&dvd&78.80&19.23&78.70&3.98& \bf80.43&23.22\\
&music&77.72&16.70&77.32&3.72& \bf78.05&21.60\\
\addlinespace
&books&73.09&15.21&71.06&1.27& \bf77.00&10.47\\
Japanese& dvd &71.10&14.86&75.75&1.48& \bf76.37&11.84\\
&music&75.15&13.72&76.22&1.83& \bf77.34&13.39\\
\addlinespace[2pt]
\hdashline[2pt/2pt]
\addlinespace[2pt]
German&-&89.69&16.19&88.73&0.92& \bf 92.61&8.38\\
French&-&87.59&16.29&89.65&1.36&\bf 90.57&11.37\\
Japanese&-&82.83&16.71&84.26&1.23&\bf 85.03&10.15\\


\hline
\addlinespace[2pt]
\end{tabular*}

\raggedright
The effect of different regularization terms on the performance of CL-SCL for cross-language sentiment (first nine rows) and topic classification (last three rows). d gives the density of the parameter matrix $\mathbf{W}$, i.e., the number of non-zero entries divided by the total number of entries.  $\mathbf{W}$ is $450 \times |V|$ where $|V|$ is in $10^5$ orders of magnitude (see Table~\ref{tab:table-dataset-statistics} for details). Elastic Net uses $\alpha=0.85$. 
\caption{Effect of regularization. }
\label{tab:table-regularization}
\end{acmtable}


\subsection{Interpretation of Results}

Primarily responsible for the effectiveness of CL-SCL is its task specificity, i.e., the way in which context contributes to meaning (pragmatics). Due to the use of task-specific, unlabeled data, relevant characteristics are captured by the pivot classifiers. 

Table~\ref{tab:table-alignments} exemplifies this claim with two pivots for German book reviews. The rows of the table show a selection of words which have the highest correlation with the pivots $\{$beautiful$_\sLang$, sch{\"o}n$_\tLang\}$ and $\{$boring$_\sLang$, langweilig$_\tLang\}$. We can distinguish between
\Ni
correlations that reflect similar meaning, such as ``amazing'', ``lovely'', or ``plain'', and 
\Nii
correlations that reflect the pivot pragmatics with respect to the task, such as ``picture'', ``poetry'', or ``pages''.

Note in this connection that the authors of book reviews tend to use the word ``beautiful'' to refer to illustrations or to poetry, and that they use the word ``pages'' to indicate lengthy or boring books. While the first type of word correlations can be obtained by methods that operate on parallel corpora, the second correlation type requires an understanding of the task-specific language use.

\begin{acmtable}{\columnwidth}
\centering
\renewcommand{\tabcolsep}{8pt}

\begin{tabular}{@{}p{0.12\columnwidth}@{\quad} p{0.15\columnwidth}@{\quad}p{0.15\columnwidth} p{0.22\columnwidth}@{\quad}p{0.22\columnwidth}@{}}
\raisebox{-2ex}{\bfseries Pivot} & \multicolumn{2}{c}{\bfseries English} & \multicolumn{2}{c}{\bfseries German} \\[-1.5ex]
\cmidrule(rr){2-3} \cmidrule(ll){4-5} 
& \multicolumn{1}{c}{Semantics} & \multicolumn{1}{c}{Pragmatics} & \multicolumn{1}{c}{Semantics} & \multicolumn{1}{c}{Pragmatics} \\
\hline
$\{$beautiful$_\sLang$, sch{\"o}n$_\tLang\}$ &
amazing, beauty, lovely & picture, pattern, poetry, photographs, paintings&
sch{\"o}ner (more beautiful), traurig (sad) & bilder (pictures), illustriert (illustrated)
\\
\addlinespace
\hdashline[2pt/2pt]
\addlinespace
$\{$boring$_\sLang$, langweilig$_\tLang\}$&
plain, asleep, dry, long & characters, pages, story&
langatmig (lengthy),  einfach (plain), entt{\"a}uscht (disappointed) & charaktere (characters), handlung (plot), seiten~(pages)
\\
\hline
\addlinespace[2pt]
\end{tabular}

\raggedright
 Semantic and pragmatic correlations identified for the two pivots  $\{$beautiful$_\sLang$, sch{\"o}n$_\tLang\}$ and $\{$boring$_\sLang$, langweilig$_\tLang\}$ in English and German book reviews.
\caption{Semantic and pragmatic correlations.}
\label{tab:table-alignments}
\end{acmtable}


\section{Conclusions}
\label{sec:conclusions}

We have presented Cross-Language Structural Correspondence Learning, CL-SCL, as an effective technology for cross-lingual adaptation. CL-SCL builds on Structural Correspondence Learning, a recently proposed algorithm for domain adaptation in natural language processing. CL-SCL uses unlabeled documents along with a feature translation oracle to automatically induce task-specific, cross-lingual feature correspondences.

We evaluated the approach for cross-language text classification, a special case of cross-lingual adaptation. The analysis covers performance and sensitivity issues in the context of sentiment and topic classification with English as source language and German, French, and Japanese as target languages. The results show a significant improvement of the proposed approach over a machine translation baseline, reducing the relative error due to cross-lingual adaptation by an average of 59\% (sentiment classification) and 30\% (topic classification) over the baseline.

Furthermore, the Elastic Net is proposed as an effective means to obtain a sparse parameter matrix, again leading to a significant improvement upon previously reported results. Note Elastic Net has implications beyond CL-SCL, in particular for Structural Correspondence Learning \cite{Blitzer:2006} and Alternating Structural Optimization \cite{Ando:2005b}.



\end{document}